\documentclass[superscriptaddress,nofootinbib,showpacs]{revtex4}
\usepackage{latexsym,epsfig,graphicx,amssymb}
\usepackage{amsmath}
\usepackage{color}%

\newcommand{\be}{\begin{eqnarray}}
\newcommand{\ee}{\end{eqnarray}}
\newcommand{\bea}{\begin{eqnarray}}
\newcommand{\nn}{\nonumber}
\newcommand{\eea}{\end{eqnarray}}

\def\k{\kappa}

\newcommand{\fr}{\frac}
\def\de{\partial}

\begin{document}

\title{Building a Holographic Superconductor with  a  Scalar Field Coupled Kinematically  to Einstein Tensor}

\author{Xiao-Mei Kuang}
\email{xmeikuang@gmail.com} \affiliation{Instituto de
F\'{i}sica, Pontificia Universidad Cat\'olica de Valpara\'{i}so,
Casilla 4059, Valpara\'{i}so, Chile.}

\author{Eleftherios Papantonopoulos}
\email{lpapa@central.ntua.gr} \affiliation{Physics Division,
National Technical University of Athens, 15780 Zografou Campus,
Athens, Greece.}

\date{\today}

\begin{abstract}

We study the holographic dual description of a superconductor in which the gravity sector consists of a Maxwell field and a charged scalar
 field  which
 except its minimal coupling to gravity it is also coupled  kinematically to Einstein tensor. As the strength of the new coupling is increased,
 the critical temperature below which the scalar field condenses is lowering, the condensation gap decreases faster than the temperature, the width of the condensation gap is not proportional to the size of the condensate and
at low temperatures the condensation gap tends to zero  for the strong coupling. These effects which are the result of the presence of the coupling of the scalar field to the Einstein tensor in the gravity bulk, provide a dual description of impurities concentration in   a superconducting state on the boundary.

\end{abstract}
\pacs{11.25.Tq, 04.70.Bw, 74.20.-z}\maketitle

\maketitle
\section{Introduction}

Employing the AdS/CFT correspondence it was shown \cite{Hartnoll:2008vx} that a
holographic superconductor can be builded on a boundary of
a gravity theory consisting of a black hole which acquires scalar hair
at temperature below a critical temperature, while above the critical temperature there is
no scalar hair. A condensate of the charged
scalar field is formed through its coupling to a Maxwell field of
the gravity sector. The study was carried out in the probe limit~\cite{Gubser:2008px}, in which in general the product of the charge of the black hole
and the charge of the scalar field is held fixed while the latter
is taken to infinity and this resulted to neither field to backreact on the metric.  Considering fluctuations of the vector
potential, the frequency dependent conductivity was calculated,
and it was shown that it develops a gap determined by the
condensate.

This model was further studied  \cite{Hartnoll:2008kx}
beyond the probe limit. If the scalar charge is large then
the  backreaction of the scalar field to the spacetime metric has
to be considered. It was found that all the
basic characteristics of the dual superconductor were
persisting even  for very small charge.  One of the characteristics of these models is
the strong pairing mechanism which is in operation. This is due to strong bounding of Cooper pairs and
this is manifest to the high values of the gap at low temperature. This strong pairing mechanism resulted to $2 \Delta \approx 8.4 T_c$, where $\Delta$ is the condensation gap, which has to be compared to
 the BCS prediction  $2 \Delta \approx 3.54 T_c$, which is much lower in real materials, due to impurities. The above proposals  have inspired much effort on  further research and on their possible extensions (see \cite{Horowitz:2010gk,Cai:2015cya} and references therein for a review).

The effects of  paramagnetic impurities on
superconductors were studied  in \cite{Abrikosov}.
It was found  that the transition
temperature decreases sharply with increasing impurity
concentration and goes to zero at a critical concentration.
 Measurements
of the energy gap and of the transition temperature
$T_c$ as a function of the concentration of paramagnetic
impurities  show that the gap decreases much more
rapidly than does the critical temperature which agrees with the results
of \cite{Abrikosov}. This behaviour was also observed in \cite{Gennes} and
 \cite{Phillips}.

Further  calculations on the effects of paramagnetic impurities in superconductors were carried out in \cite{skalski},
 taking full advantage of the information contained
in the Green's function of the system. The density of states in energy has been computed for different
values of the inverse collision time. The  excitation energy gap  $\Omega_g$  is defined to be the energy at which the density of states vanishes.
The temperature-dependent order parameter has been computed and the
behaviour of $\Omega_g (T)$ was determined. A comparison with tunneling experiments showed a disagreement of the two parameters.
 The real part of the conductivity at $T=0$ is shown to be zero for frequencies less
than $2 \Omega_g$ and proportional to the square of the density of states for vanishingly small frequencies in the gapless
region of impurity concentrations. These detailed calculations showed clearly that in real materials, the essential feature of superconductivity is the correlation of the
electrons and the formation of Cooper pairs, rather than the existence of a gap in the excitation energy spectrum. For even in the absence of a gap, so
long as correlations persist there is an ordered state below a certain critical temperature which displays the usual
properties of a superconductor.

A holographic realization of impurities was discussed in \cite{Hartnoll:2008hs}.  Using  the  AdS/CFT  correspondence
two  different  types  of  impurities were calculated and the Nernst response for the impure theory was specified.
A model of a gravity dual of a gapless
superconductor was proposed in \cite{Koutsoumbas:2009pa}. The gravity sector was consisting of a charged scalar field
which provided
the scalar hair of an exact black hole
solution~\cite{Martinez:2004nb,Martinez:2005di} below a critical temperature $T_c$, while  an electromagnetic perturbation  of the background
determined the conductivity giving rise to a gapless superconductor. It was found that the  normal component
of the DC conductivity  had a milder behaviour than
the dual superconductor in the case of a black hole of flat
horizon~\cite{Hartnoll:2008vx} in which $n_{n}$ exhibited a clear
gap behaviour. The  behaviour it was observed in  the boundary
conducting theory was attributed  to materials with paramagnetic
impurities as it was discussed in \cite{skalski}. Impurities effects were studied in \cite{Ishii:2012hw}, while impurities in the Kondo
 Model were considered in \cite{O'Bannon:2015gwa,Erdmenger:2015xpq}.

In this work we extend the model in \cite{Hartnoll:2008vx} by introducing a derivative coupling of the
scalar field to Einstein tensor.
This term belongs to a general class of
scalar-tensor gravity theories resulting from the Horndeski
Lagrangian \cite{Horndeski:1974wa}. These theories, which were
recently rediscovered \cite{Deffayet:2011gz},
 give second-order field equations and contain as a subset a theory
which preserves classical Galilean symmetry \cite{Nicolis:2008in,
Deffayet:2009wt, Deffayet:2009mn}. The derivative coupling of the scalar field to Einstein tensor
introduces a new scale in the theory which on short distances allows
to find  black hole solutions \cite{Kolyvaris:2011fk,Rinaldi:2012vy,Kolyvaris:2013zfa,Cisterna:2014nua,Brihaye:2016lin} with scalar hair just outside the black hole horizon,
while if one considers the gravitational collapse of a scalar field coupled to the Einstein tensor then the formation of a black hole
takes more time to be formed compared to the collapse of a scalar field minimally coupled to gravity  \cite{Koutsoumbas:2015ekk}.
On large distances  the presence of the derivative coupling acts as a friction term in the inflationary period of the cosmological
evolution  \cite{Amendola:1993uh,Sushkov:2009hk,germani}. Moreover, it was found that at the end of
inflation in the preheading period, there is a suppression of heavy particle production  as the derivative coupling is increased. This was attributed to the fast decrease of  kinetic
energy of the scalar field due to its  wild oscillations \cite{Koutsoumbas:2013boa}.

The above discussion indicates that one of the main effects of the kinematic coupling of a scalar field to Einstein tensor
is that  gravity  influences strongly  the  propagation of the scalar field compared to a scalar field minimally coupled to gravity.
 We are going to use this behaviour of the scalar field
to holographically simulate the effects of a high concentration of impurities in a material.  The presence of impurities in a superconductor is
making the pairing mechanism of forming Cooper pairs less effective and this is happening because the quasiparticles are loosing energy because of strong
concentration of impurities. This holographic correspondence is supported by our finding. We found that as the value of the derivative coupling is increased the critical temperature is degreasing while
the condensation gap $\Delta$  is decreasing faster than the temperature. Also by calculating the perturbation of the scalar potential we found
that the condensation gap for large values of the derivative coupling is not proportional  to the frequency of the real part of conductivity
which is characteristic of a superconducting state with impurities \footnote { We note that in \cite{Lin:2015kjk} a holographic
superconductor was considered  containing a  derivative coupling
for a scalar field in the background of a regular phantom plane symmetric black hole.}.

The work is organized as follows. In Section \ref{review} we review the basic ingredients of the holographic model discussed in \cite{Hartnoll:2008vx}.
In Section \ref{derscalar} we extend the holographic model of  Section \ref{review} adding a scalar field coupled to Einstein tensor and we give the equations of motion. In Section \ref{solution} we solve numerically the equations of motion and we find the condensation gap and the critical temperature for various values of the derivative coupling. In Section \ref{conductivity} we study the conductivity while in Section \ref{conc} are our conclusions.

\section{Holographic Superconductor with a scalar field minimally coupled to gravity}
\label{review}

In this section we review in brief the holographic model of \cite{Hartnoll:2008vx}.
To introduce the minimal holographic superconductor model, we  consider a Maxwell field and a charged complex scalar field
coupled to gravity with the action
\begin{eqnarray}\label{System}
S=\int d^{4}x\sqrt{-g}\left[\frac{R+\Lambda}{16 \pi G}
-\frac{1}{4}F_{\mu\nu}F^{\mu\nu}-|\nabla\psi - iqA\psi|^{2}
-m^2|\psi|^2 \right] \ .
\end{eqnarray}
If we rescale the fields and the coupling constant as \bea
 A_{\mu} =\frac{\tilde{A}_{\mu}}{q}~,~~~
 \psi =\frac{\tilde{\psi}}{q}~,
 \eea
then the matter action in (\ref{System}) has a $1/q^2$ in front,
so the backreaction of the matter fields on the metric is
suppressed when $q$ is large and the  limit $q\rightarrow\infty$ defines the probe limit.
In this limit
the Einstein equations
admit as a solution the planar Schwarzschild AdS black hole
 \be\label{schwar}
 ds^2= -f(r) dt^2 + \frac{dr^2}{f(r)} + r^2 (dx^2 +
 dy^2)\qquad,\qquad
f(r)=r^2 - \frac{r_h^3}{r}~. \ee

 Taking the ansatz
$\psi=|\psi|$, $A=\phi dt$ where $\psi$, $\phi$ are both functions
of $r$ only, we can obtain the equations of motion for $\psi,
\phi$
\begin{eqnarray}
\psi^{\prime\prime}+\left(
\frac{f^\prime}{f}+\frac{d-2}{r}\right)\psi^\prime
+\left(\frac{\phi^2}{f^2}-\frac{m^2}{f}\right)\psi=0\,,
\label{Psi}
\end{eqnarray}
\begin{eqnarray}
\phi^{\prime\prime}+\frac{d-2}{r}\phi^\prime-\frac{2\psi^2}{f}\phi=0~.
\label{Phi}
\end{eqnarray}
It was argued in \cite{Gubser:2008px} that there is a critical temperature
below which the
 black
hole  acquires hair~\cite{Hartnoll:2008vx,Hartnoll:2008kx}.

The equations (\ref{Psi})  and (\ref{Phi}) can be solved
numerically by doing integration from the horizon out to the
infinity taking under consideration the right boundary conditions.
The solutions behave like
\begin{eqnarray}
\psi=\frac{\psi_{-}}{r^{\lambda_{-}}}+\frac{\psi_{+}}{r^{\lambda_{+}}}\,,\hspace{0.5cm}
\phi=\mu-\frac{\rho}{r}\,, \label{infinity}
\end{eqnarray}
with
\begin{eqnarray}
\lambda_\pm=\frac{1}{2}\Big{[}3 \pm\sqrt{(9+4m^{2}L
^2)}~\Big{]}\,,\label{LambdaZF}
\end{eqnarray}
where $L$ is the  scale of the AdS space and $\mu$ and $\rho$ are
interpreted as the chemical potential and charge density in the
dual field theory respectively. The coefficients $\psi_{-}$ and
$\psi_{+}$  according to the AdS/CFT correspondence, 
correspond to the vacuum expectation values
$\psi_{-}=\langle\mathcal{O}_{-}\rangle/\sqrt{2}$, $\psi_{+}=\langle\mathcal{O}_{+}\rangle/\sqrt{2}$ of an
operator $\mathcal{O}$ dual to the scalar field. We can impose
boundary conditions that either $\psi_{-}$ or $\psi_{+}$ vanishes.

 Fluctuations of the vector
potential $A_x$ in the gravity sector gives the
 conductivity in the dual CFT as a function of
frequency. The Maxwell equation for
zero spatial momentum and with a time dependence of the form $e^{-
i \omega t}$ reads as
 \be\label{eq:Axeq} A_x'' + \frac{f'}{f} A_x' +
\left(\frac{\omega^2}{f^2} - \frac{2 \psi^2}{f} \right) A_x = 0
\,. \ee
The above equation can be solved by imposing ingoing wave boundary conditions at the horizon
 $A_x  \propto f^{-i\omega/3r_0}$.  On the other hand, the
asymptotic behaviour of the Maxwell field at boundary  is
 \be A_x = A_x^{(0)} + \frac{A_x^{(1)}}{r} + \cdots. \ee
 Then according to  AdS/CFT correspondence  dual source and expectation
value for the current are given by \be A_x = A_x^{(0)} \,, \qquad
\langle J_x \rangle = A_x^{(1)} \,. \ee
Finally, Ohm's law gives  the conductivity \be\label{eq:conductivity} \sigma(\omega)
= \frac{\langle J_x \rangle}{E_x} = - \frac{ \langle J_x
\rangle}{\dot A_x} = -\frac{ i \langle J_x \rangle}{\omega A_x}
= - \frac{i A_x^{(1)}}{\omega A_x^{(0)}} \,. \ee

\section{Holographic Superconductor with a scalar field kinematically coupled to Einstein Tensor}
\label{derscalar}

In this section, we will consider a complex scalar field which excepts its coupling
to curvature it is also coupled to Einstein tensor with action
\bea\label{EGBscalaraction}
   I=\int  d^4x\sqrt{-g}\left[ \frac{R+\Lambda}{16\pi G}-\fr{1}{4}F_{\mu\nu}F^{\mu\nu}-(g^{\mu\nu}+\k
   G^{\mu\nu})D_\mu\psi (D_\nu\psi)^{*} - m^2|\psi|^2 \right]~,
\eea
where \be D_\mu = \nabla_\mu - i e A_\mu \ee
and $e$, $m$ are the charge and the mass of the scalar field and
$\k$ the coupling of the scalar field to Einstein tensor of
dimension length squared. For convenience we set
 \bea
\Phi_{\mu\nu} &\equiv& D_{\mu}\psi (D_{\nu}\psi)^*~,\\
\Phi &\equiv& g^{\mu\nu}\Phi_{\mu\nu}~,\\
C^{\mu\nu} &\equiv& g^{\mu\nu} + \k G^{\mu\nu}~.
 \eea
The field equations resulting from the action
(\ref{EGBscalaraction}) are
 \bea G_{\mu\nu} +\Lambda g_{\mu\nu} = 8\pi  T_{\mu\nu}
\ , \ \ \ \ T_{\mu\nu} = T_{\mu\nu}^{(\psi)} +
T_{\mu\nu}^{(EM)} + \k\Theta_{\mu\nu}~, \label{einst}\eea where,
\bea
T_{\mu\nu}^{(\psi)} & = &   \Phi_{\mu\nu} + \Phi_{\nu\mu} - g_{\mu\nu}(g^{ab}\Phi_{ab} + m^2 |\psi|^2)~, \\
T_{\mu\nu}^{(EM)} & = & F_{\mu}^{\phantom{\mu} \alpha} F_{\nu
\alpha} - \fr{1}{4} g_{\mu\nu} F_{\alpha\beta}F^{\alpha\beta}~,
\eea and \bea
\Theta_{\mu\nu}  = & -& g_{\mu\nu} R^{ab}\Phi_{ab} + R_{\nu}^{\phantom{\nu}a}(\Phi_{\mu a} + \Phi_{a\mu}) + R_{\mu}^{\phantom{\mu}a} (\Phi_{a\nu} + \Phi_{\nu a})  - \fr{1}{2} R (\Phi_{\mu\nu} + \Phi_{\nu\mu}) \nn\\
& - & G_{\mu\nu}\Phi - \fr{1}{2}\nabla^a\nabla_\mu(\Phi_{a\nu} + \Phi_{\nu a}) - \fr{1}{2}\nabla^a\nabla_\nu(\Phi_{\mu a} + \Phi_{a\mu}) + \fr{1}{2}\Box (\Phi_{\mu\nu} + \Phi_{\mu\nu}) \nn \\
& + & \fr{1}{2}g_{\mu\nu} \nabla_a\nabla_b (\Phi^{ab} + \Phi^{ba})
+ \fr{1}{2}(\nabla_\mu\nabla_\nu + \nabla_\nu\nabla_\mu) \Phi -
g_{\mu\nu}\Box\Phi~. \label{theta} \eea
The Klein-Gordon equation is
\be (\de_\mu-i e A_\mu) \left[
\sqrt{-g}C^{\mu\nu}(\de_\nu - i e A_\nu)\psi\right] =
\sqrt{-g} m^2\psi~, \label{glgord} \ee
while the Maxwell equations read \be \nabla_\nu F^{\mu\nu} +
C^{\mu\nu} \left[ 2 e^2 A_\nu |\psi|^2 + i e
(\psi^*\nabla_\nu\psi- \psi\nabla_\nu\psi^*)\right]
=0~. \label{max}\ee

\subsection{Solution of equations of motion and phase transition}
\label{solution}

As in \cite{Hartnoll:2008vx} we also take the ansatz
$\psi=|\psi|$, $A=\phi dt$ where $\psi$, $\phi$ are both functions
of $r$ only. In the probe limit, under the metric (\ref{schwar}), the above equations of motion for the matter fields become

\begin{equation}
\left[1+\kappa\left(\frac{f}{r^2}+\frac{f'}{r}\right)\right]\psi^{\prime\prime} + \left[\frac{2}{r} + \frac{f^\prime}{f} + \kappa\left (\frac{3f'}{r^2}+\frac{f'^{2}}{r f}+\frac{f''}{r} \right)\right]\psi^\prime + \left[\frac{e^2\phi^2}{f^2}\left(1+\kappa\left(\frac{f}{r^2}+\frac{f'}{r}\right)\right) - \frac{m^2}{f}\right]\psi = 0 \label{newpsi}~,
\end{equation}

\begin{equation} \phi^{\prime\prime} + \frac{2}{r}\phi^\prime - \frac{2e^2\psi^2}{f}\left[1+\kappa\left(\frac{f}{r^2}+\frac{f'}{r}\right)\right]\phi = 0~. \label{newphi}
\end{equation}

Note that if $\kappa=0$ then we get the equations  (\ref{Psi}) and (\ref{Phi}).
The new equations depend now  on the second derivative of the function $f$ in contrast to the preview case where we had only its first derivative, due to the additional term of $G^{\mu\nu}$ in the action (\ref{EGBscalaraction}).

With the same study in the minimal case presented in previous section, we can numerically solve the above equations
of motion.  We especially study the effect of $\kappa$ on the critical temperature and strength of condensation in the dual
supercondutor. Without loss of generality, we set $e=1$ and $m^2=-2$, so that we have $\lambda_-=1$ while $\lambda_+=2$. We will choose $\psi_-$ as the source which is set to be vanishing, while $\psi_+$ as vacuum expectation values.
\begin{table}
  \centering
  \begin{tabular}{|c||c|c|c|c|c|c|c|}
    \hline
    $\kappa$&-0.01&0 & 0.01 &0.05 &  0.1&0.5 &1 \\ \hline
    $\frac{T_c}{\sqrt{\rho}}$&$0.1218$ & $0.1184$&$0.1158$&$0.1091$&$0.1043$&$0.09353$&$0.0906$ \\ \hline
    $C_1$&$243$ & $140$&$73$&$10$&$2$&$0.01$&$0.003$ \\ \hline
  \end{tabular}
  \caption{\label{tableTc}Critical temperature $T_{c}$ for the phase transition of $<O_{+}>$ with different $\kappa$.}
\end{table}
\begin{figure}
  \centering
  \includegraphics[width=.4\textwidth]{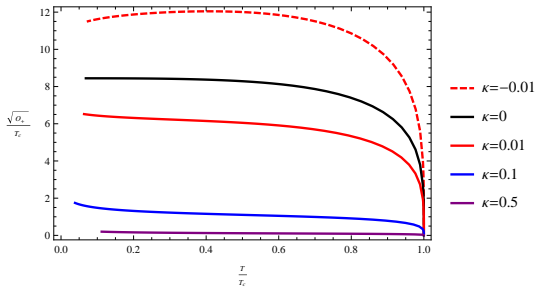}\hspace{1cm}
  \caption{\label{fig-condensation}The strength of condensation VS the temperature with samples of $\kappa$.}
\end{figure}

The critical temperatures with different $\kappa$ are summarized in Table {\ref{tableTc}}. It shows that as the coupling becomes stronger, the critical temperature is lower, meaning that the phase transition is hard to occur as it is expected. This effect of coupling is very different from the influence of higher-derivative coupling between the matters studied in \cite{Kuang:2013oqa,Kuang:2014kha}. As the temperature is decreased, the scalar field can have non-zero solution and the strength of condensation becomes higher. We show this
phenomena in FIG. \ref{fig-condensation}. It is obvious that the strength of condensation is suppressed by the derivative coupling. And with enlarging the coupling\footnote{There is an instability in our numerical solution of the equations of motion for large or small values of the derivative coupling when the temperature is far way from $T_c$. Specially, it is difficult to reach a low temperature $T/T_c\sim 0.1$ when $\kappa>0.5$.  For this reason in FIG. \ref{fig-condensation}  we restricted the curves to the values of $\kappa$ in the range $ -0.01\leqslant\kappa\leqslant0.5$. However, when we are near $T_c$ our numerics work better, so we could fit relation (\ref{near}) for larger values of $\kappa$. So in order to study the same shift from the critical temperature   in the following  study we focus on the range $ -0.01\leqslant\kappa\leqslant0.5$.      }, the condensation decreases faster than the temperature, so that the condensation gap tends to be zero at low temperatures for the strong coupling. This behaviour from $\kappa$ is coincident with the effect of paramagnetic impurities on superconductors observed in \cite{Abrikosov,Gennes,Phillips}.

We fit the curve near the critical temperatures. As $T\rightarrow T_c$, the condensation is continuous and  behaves as
\begin{equation}
\langle\mathcal{O}_{+}\rangle\simeq C_1 T_c^2(1-T/T_c)^{1/2} \label{near}
\end{equation}
where $C_1$ are also listed in Table {\ref{tableTc}}. We see that the coefficient $C_1$ decreases drastically to suppress the condensation when the coupling is increased. Observe that the exponent is always $1/2$ which implies that $\kappa$ does not modify the order of the phase transition and it is always second order. In the next subsection  we  will study the conductivity caused by the condensation below the critical temperature.

\subsection{Conductivity}
\label{conductivity}

To compute the conductivity in the dual CFT as a function of
frequency we need to solve the Maxwell equation for fluctuations of the vector
potential $A_x$.   The Maxwell equation at
zero spatial momentum and with a time dependence of the form $e^{-i \omega t}$ gives
\begin{equation}\label{eq:Axeq}
A_x'' + \frac{f'}{f} A_x' +\left[\frac{\omega^2}{f^2} - \frac{2 e^2\Psi^2}{f}
\left(1+\kappa\left(\frac{f''}{2}+\frac{f'}{r}\right)\right)\right] A_x = 0~.
\end{equation}

We will solve the perturbed Maxwell equation with ingoing wave boundary conditions at the horizon, i.e.,
 $A_x  \propto f^{-i\omega/3r_0}$. The asymptotic behaviour of the Maxwell field at large radius  is  $A_x = A_x^{(0)} + \frac{A_x^{(1)}}{r} + \cdots$. Then,
according to AdS/CFT dictionary,  the dual source and expectation
value for the current are given by $ A_x = A_x^{(0)} $ and $ \langle J_x \rangle = A_x^{(1)}$, respectively.
Thus, similar to the equation (\ref{eq:conductivity}), the conductivity is also read as
\begin{equation}
\sigma(\omega)= - \frac{i A_x^{(1)}}{\omega A_x^{(0)}}~.
\end{equation}
\begin{figure}[h]
  \centering
  \includegraphics[width=0.9\textwidth]{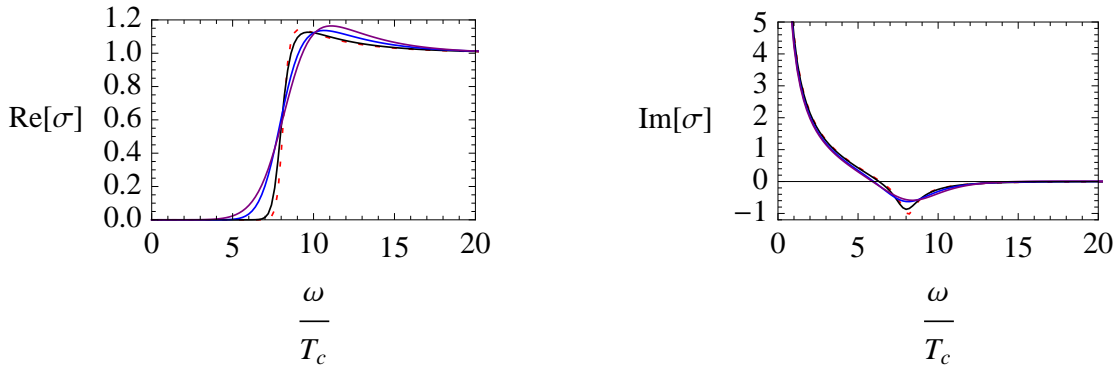}\hspace{1cm}
  \caption{\label{fig-sigma}Conductivity for samples of $\kappa$ at $T/T_c\simeq 0.1$. The values of $\kappa$ are $\kappa=-0.01$(dotted red),$\kappa=0$(black),$\kappa=0.1$(blue) and $\kappa=0.5$(purple).}
\end{figure}

The numerical results of conductivity at low temperature
with $\frac{T}{T_c}\simeq 0.1$ are shown  in FIG. \ref{fig-sigma}.  Similarly, due to the Kramers-Kronig(KK) relations
\begin{equation}\label{eq:KK}
\mathrm{Im}[\sigma(\omega)]=-\frac{1}{\pi}\mathcal{P}\int_{-\infty}^{\infty}\frac{\mathrm{Re}[\sigma(\omega')]}{\omega'-\omega}d\omega'~,
\end{equation}
where $\mathcal{P}$ denotes the Cauchy principal value, the divergence of the imaginary part at zero frequency indicates a delta function of real part at $\omega=0$.
At very low temperature, the energy gap in the real part of the conductivity measured by the critical temperature (left part of FIG. \ref{fig-sigma}) decreases as the derivative coupling becomes stronger, which agrees well with the behaviour of the condensation as it is depicted in FIG. \ref{fig-condensation}  in the last subsection.

\begin{figure}[h]
  \centering
  \includegraphics[width=0.8\textwidth]{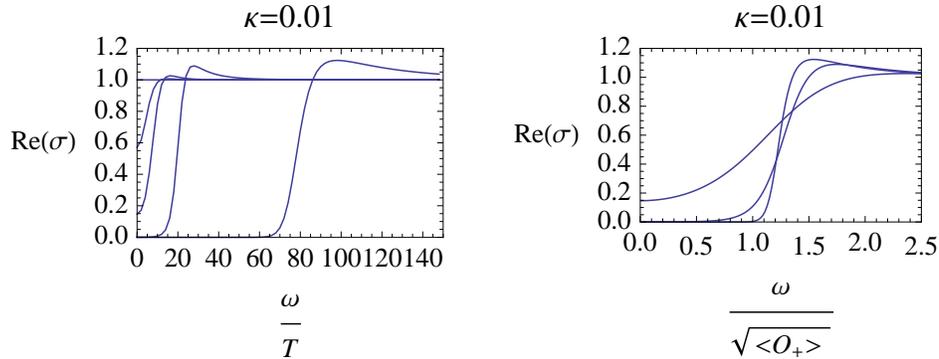}\hspace{1cm}
  \caption{\label{fig-sigmak=0.01}Conductivity with $\kappa=0.01$ as the temperature goes further away from the critical point. In the left plot, from left to right, the temperatures are $T/T_c= 1$, $T/T_c\simeq 0.94$, $T/T_c\simeq 0.80$, $T/T_c\simeq 0.40$, $T/T_c\simeq 0.10$. They are $T/T_c\simeq 0.80$, $T/T_c\simeq 0.40$, $T/T_c\simeq 20$ in the right plot.}
\end{figure}

In order to show how the conductivity opens a gap as the temperature decreases, in FIG. \ref{fig-sigmak=0.01}, we show the frequency-dependent conductivity with $\kappa=0.01$ at different temperatures. In the left plot, the horizontal line, which is frequency independent, corresponds to temperatures at or above the critical value and  there is no condensate. As we lower the temperature in the subsequent curves, we see that a gap $\omega_g$ opens in the real part of the conductivity and the gap becomes wider and deeper as the temperature becomes smaller. In the right plot, we plot the conductivity for temperature lower than
$T_c$ related the left plot by rescaling the frequency by the condensate. It is obvious that the curves tend to a limit in which the width of the gap is proportional to the size of the condensate. i.e., $\omega_g \simeq \sqrt{<O_+>}$. The above features are similar with those in minimal coupling disclosed in \cite{Hartnoll:2008vx} .

\begin{figure}[h]
  \centering
  \includegraphics[width=0.8\textwidth]{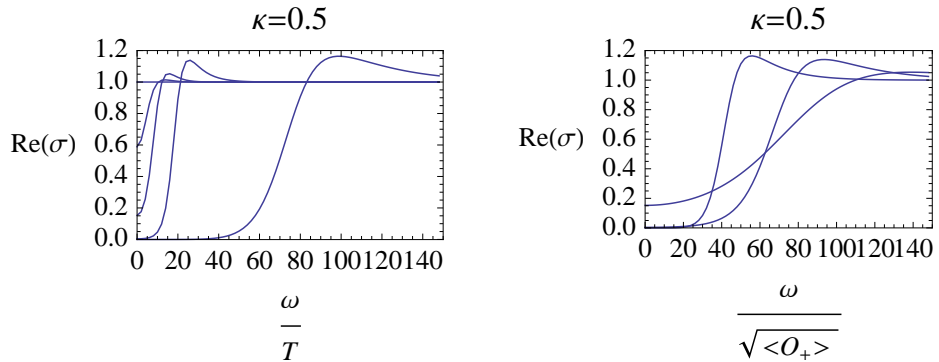}\hspace{1cm}
  \caption{\label{fig-sigmak=0.5}Conductivity for $\kappa=0.5$  with the frequency rescaled by temperature and the condensation.  In the left plot, from left to right, the temperatures are $T/T_c= 1$, $T/T_c\simeq 0.94$, $T/T_c\simeq 0.80$, $T/T_c\simeq 0.40$, $T/T_c\simeq 0.10$.  They are $T/T_c\simeq 0.80$, $T/T_c\simeq 0.40$, $T/T_c\simeq 0.10$ in the right plot.}
\end{figure}

Furthermore, the frequency-dependent conductivity with $\kappa=0.5$ at different temperatures is presented in
FIG. \ref{fig-sigmak=0.5}. Similarly to the cases with $\kappa=0.01$ and minimal coupling, lower temperature gives us wider energy gap. By comparing the left plots of FIG. \ref{fig-sigmak=0.01} and FIG. \ref{fig-sigmak=0.5},  we see that at the same scaled temperature $T/T_c$, the gap is narrower for larger coupling, which is consistent with the results shown in  FIG. \ref{fig-condensation} and FIG. \ref{fig-sigma}. However, if being scaled by the condensation, see the right plot for
$\kappa=0.5$, the behavior $\omega_g \simeq \sqrt{<O_+>}$ at low $T/T_c$ is violated and it is $\omega_g \simeq 20\sqrt{<O_+>}$ for $T/T_c\simeq 0.1$. The coefficient $20$ is not out of expect.  Because if we see carefully the condensation plot
FIG. \ref{fig-condensation} and FIG. \ref{fig-sigma}, we will find  $ \sqrt{<O_+>}\simeq 0.2 T_c$ and $\omega_g \simeq 4T_c$, so we get the same relation as before  $\omega_g \simeq 20\sqrt{<O_+>}$.
\begin{table}[h]
  \centering
  \begin{tabular}{|c||c|c|c|c|c|c|}
    \hline
    $\kappa$&-0.01&0 & 0.01 &0.05 &  0.1&0.5  \\ \hline
    $\frac{\omega_g}{\sqrt{\langle O_+\rangle}}$&$0.61$ & $0.80$&$1$&$1.9$&$2.7$&$20$ \\ \hline
  \end{tabular}
  \caption{\label{tablewg}$\frac{\omega_g}{\sqrt{\langle O_+\rangle}}$ far away from the critical temperature ($T/T_c\simeq 0.10$) with different $\kappa$.}
\end{table}

For the systems far away from the critical temperature ($T/T_c\simeq 0.10$), by careful study,
we summarize the values of $\frac{\omega_g}{\sqrt{<O_+>}}$ with some
$\kappa$ in Table \ref{tablewg}. It is obvious that as $|\kappa|$ shifts from the minimal coupling, the relation
$\omega_g \simeq \sqrt{<O_+>}$ is sharply violated, which is one of the main features of a superconducting state with impurities. Note that this effect of possible impurity parameter $\kappa$ is different from that studied in \cite{Ishii:2012hw} where it can not introduce this kind of violation.

We can now  explore the behaviour of conductivity at very low frequency. When $T<T_c$, the real part of the conductivity present a delta function at zero frequency and the imaginary part has a pole, which is attributed to the KK relations (\ref{eq:KK}).
More specifically, as $\omega\rightarrow 0$, the imaginary part behaves as $Im(\sigma)\sim n_s/\omega$, and according to Kramers-Kronig relations, the real part has the form $Re(\sigma)\sim\pi n_s\delta(\omega)$. Here the coefficient $n_s$ of the delta function is defined as the superfluid density. By fitting data near the critical temperature, we find that with various couplings, the superfluid density has the behaviour
\begin{equation}\label{eq:ns}
n_s\simeq C_2 T_c(1-T/T_c)~,
\end{equation}
which means that $n_s$ will vanish linearly as $T$ goes to $T_c$. This is consistent with that happens in the minimal coupling. Also we find that the coefficient $C_2$ is not very sensitive to the coupling and the value oscillates around $24$ with $\pm 2$ shifting for the samples of $\kappa$. This  insensitivity to the changes of the coupling is reasonable because in the region very close to the critical temperature, the system is marginally shifted from the normal state, so that varying $\kappa$ has very small effect on the behaviour of the solutions.  Meanwhile, as proposed in \cite{Hartnoll:2008vx}, the non-superconducting density can be defined as
$n_n=\lim_{\omega\rightarrow 0}Re(\sigma)$. Far from the critical temperature (with $T/T_c\sim 0.1$), we fit and obtain that $n_n$ decays as
\begin{equation}\label{eq:nn}
n_n\sim \exp[-\frac{\Omega_g}{T}]~,
\end{equation}
where $\Omega_g$ can be explained as the energy gap for charged excitation at the corresponding temperature. The values of
$\Omega_g$ close to $T/T_c\sim 0.1$ are summarized in Table \ref{tableWg}. Rough comparison with the numerical energy gap
$\omega_g/T_c$ shown in the left plot of FIG.\ref{fig-sigma} gives us that $\omega_g \simeq 2\Omega_g$, where the factor $2$ suggests that the gaped charged quasiparticles  are formed in pairs as addressed in \cite{Hartnoll:2008vx}. Analyzing the data in the table, we see that $\kappa=0.5$ corresponds to $\omega_g \simeq 2\Omega_g \simeq 4 T_c$ which is very near the prediction  $\omega_g \simeq 3.54 T_c$ in BCS theory. In this tendency, we believe that  BCS prediction may be fulfilled by larger coupling once the numerics are  controlable.  Thus, in this sense, the derivative coupling somehow mimic the effect of the impurities in a real material.

\begin{table}[h]
  \centering
  \begin{tabular}{|c||c|c|c|c|c|c|}
    \hline
    $\kappa$&-0.01&0 & 0.01 &0.05 &  0.1&0.5 \\ \hline
    $\frac{\Omega_g}{T_c}$&$4.0$ & $3.8$&$3.5$&$3.0$&$2.8$&$2.0$\\ \hline
  \end{tabular}
  \caption{\label{tableWg} The energy gap for charged excitation $\frac{\Omega_g}{T_c}$ with different $\kappa$.}
\end{table}

\section{Conclusions}
\label{conc}

We studied a holographic description of a superconductor in which the gravity sector consists of a Maxwell field and a charged scalar field which except its usual minimal coupling to gravity it is also coupled to Einstein tensor. Solving the equations of motion numerically in the probe limit, we found that as the strength of the new coupling is increased,
 the critical temperature below which the scalar field condenses is lowering, the condensation gap decreases faster than the temperature, the width of the condensation gap is not proportional to the size of the condensate and
at low temperatures the condensation gap tends to zero  for the strong coupling.
Analysing the frequency dependence of conductivity we found that at strong coupling the relation $\omega_g \simeq 2\Omega_g \simeq 4 T_c$ holds, where $\Omega_g$ is the energy gap for charged excitation, which is very near the prediction  $\omega_g \simeq 3.54 T_c$ in BCS theory.

We argued that these results suggest that the  derivative coupling in the gravity bulk can have a dual interpretation on the boundary corresponding to impurities concentrations in a real material. This correspondence can be understood from the fact that the coupling of a scalar field to Einstein tensor  alters the kinematical state of the scalar field a behaviour which it is also exhibited by  the quasiparticles moving in a material with impurities.

We believe that the probe limit captures all the essential features of the problem under study. Nevertheless, it would be interesting to extend this study beyond the probe limit. Assuming a spherically symmetric ansatz for the metric the field equations
(\ref{einst}), (\ref{glgord}) and (\ref{max}) have to be solved. However, even  their numerical solution is a formidable task mainly because of the presence of the energy-momentum tensor (\ref{theta}) resulting from the coupling of the scalar field to  Einstein tensor. A more realistic approach would be to follow
the perturbative  methods employed in \cite{Kolyvaris:2011fk,Kolyvaris:2013zfa}.

\begin{acknowledgments}

We thank Olivera Miskovic and George Filios for their contribution at the early stage of this work. X-M.K  is  partly supported by FONDECYT Grant No.3150006.

\end{acknowledgments}


\end{document}